\documentclass[prl,aps,twocolumn,superscriptaddress]{revtex4}

\usepackage[dvips]{graphicx}
\usepackage{amssymb,amsfonts,amsmath}
\usepackage{color}
\usepackage{ulem}

\newcommand{\rv}{{\mathbf r}}
\newcommand{\ev}{{\mathbf e}}

\newcommand{\Jv}{{\bf J}}

\newcommand{\fv}{{\bf f}}

\newcommand{\vel}{{\bf v}}

\newcommand{\rnt}{{\bf r}^N\!\!,t}

\begin{document}

\title{Structural nonequilibrium forces in driven colloidal systems}

\author{Nico C.\ X.\ Stuhlm\"uller}
\affiliation{Theoretische Physik II, Physikalisches Institut, 
  Universit{\"a}t Bayreuth, D-95440 Bayreuth, Germany}

\author{Tobias Eckert}
\affiliation{Theoretische Physik II, Physikalisches Institut, 
  Universit{\"a}t Bayreuth, D-95440 Bayreuth, Germany}

\author{Daniel de las Heras}
\affiliation{Theoretische Physik II, Physikalisches Institut, 
  Universit{\"a}t Bayreuth, D-95440 Bayreuth, Germany}

\author{Matthias Schmidt}
\affiliation{Theoretische Physik II, Physikalisches Institut, 
  Universit{\"a}t Bayreuth, D-95440 Bayreuth, Germany}

\date{22 December 2018; revised version: 18 July 2018}

\begin{abstract}
We identify a structural one-body force field that sustains spatial
inhomogeneities in nonequilibrium overdamped Brownian many-body
systems. The structural force is perpendicular to the local flow
direction, it is free of viscous dissipation, it is microscopically
resolved in both space and and time, and it can stabilize density
gradients. From the time evolution in the exact (Smoluchowski)
low-density limit, Brownian dynamics simulations and a novel power
functional approximation, we obtain a quantitative understanding of
viscous and structural forces, including memory and shear migration.
\end{abstract}

\maketitle

It is a very significant challenge of Statistical Physics to
rationalize and predict nonequilibrium structure formation from a
microscopic starting point.  Primary examples include shear banding
\cite{dhont1999,dhont2003,dhont2014}, where spatial regions of
different shear rate coexist, laning transitions in oppositely driven
colloids \cite{laningLitOne,laningLitTwo}, where regions of different
flow direction occur, as well as migration effects in inhomogeneous
shear flow
\cite{braderkrueger11epl,braderkrueger11molphys,reinhardt2013,
  aerov2014,aerov2015,scacchi2016}. In computer simulations,
discriminating true steady states from slow initial transients can be
difficult~\cite{zausch2008,harrer2012}. Often the nonequilibrium
structuring effects have associated long time scales and strong
correlations~\cite{Bolhuis2015,zimmermann16}. The underlying
equilibrium phase diagram and bulk structure might already be complex
and interfere with the genuine nonequilibrium
effects~\cite{floating,empty}.

To identify commonalities of all of the above situations, we
investigate here a representative model situation.  We put forward a
systematic classification of the occurring nonequilibrium forces and
identify a structural force component, which is able to sustain
density gradients without creating dissipation. The structural force
is solely due to the interaction between the particles, and it is
hence of a nature different than that of the lift forces in
hydrodynamics.  We rationalize our findings by constructing an
explicit power functional approximation, \eqref{EQPtexcAll} below.

We restrict ourselves to overdamped Brownian dynamics and consider the
microscopically resolved position- and time-dependent one-body
density, $\rho(\rv,t)$, and one-body current, $\Jv(\rv,t)$. Both
fields are related via the (exact) continuity equation $\partial
\rho/\partial t = -\nabla\cdot\Jv$; here $\nabla$ indicates the
derivative with respect to position $\rv$.  The velocity profile is
simply the ratio $\vel(\rv,t)=\Jv(\rv,t)/\rho(\rv,t)$.

The exact time evolution,
with no hydrodynamic
interactions present, can then be expressed by the one-body force
balance equation
\begin{align}
  \gamma \vel
  &= {\bf f}_{\rm int} + {\bf f}_{\rm ext} 
   - k_BT \nabla \ln \rho,
  \label{EQofMotion}
\end{align}
where $\gamma$ is the single-particle friction constant against the
(implicit) solvent, ${\bf f}_{\rm int}(\rv,t)$ is the internal force
field, ${\bf f}_{\rm ext}(\rv,t)$ is a one-body external force field
that in general drives the system out of equilibrium, and $- k_BT
\nabla \ln \rho(\rv,t) \equiv \fv_{\rm aid}(\rv,t)$ is the adiabatic
ideal gas contribution due to the free thermal diffusion; here
$k_B$ is the Boltzmann constant and $T$ is absolute temperature.  The
internal force field $\fv_{\rm int}(\rv,t)$ arises from the
interparticle interaction potential $u(\rv^N)$, where
$\rv^N\equiv\rv_1\ldots\rv_N$ denotes the set of all position
coordinates, and it can be expressed as an average over the
interparticle one-body force density ``operator''
\begin{align}
  \hat{\bf F}_{\rm int} &= 
  -\sum_i\delta(\rv-\rv_i)\nabla_i u(\rv^N),
\end{align}
where the sum is over all particles, $\delta(\cdot)$ indicates the
Dirac distribution and $\nabla_i$ denotes the derivative with respect
to $\rv_i$.  Using the probability distribution $\Psi(\rv^N,t)$ of
finding microstate $\rv^N$ at time $t$, the average is built according
to
\begin{align}
  {\bf f}_{\rm int}(\rv,t) &= 
         \rho(\rv,t)^{-1}\int d\rv^N \Psi(\rnt)
         \hat {\bf F}_{\rm int},
         \label{EQinternalForceField}
\end{align}
where the normalization is performed using the time-dependent one-body
density, defined as the average
\begin{align}
  \rho(\rv,t) = 
  \int d\rv^N \Psi(\rnt) \hat\rho,
\end{align}
where $\hat\rho=\sum_i\delta(\rv-\rv_i)$ is the density operator.

The internal force field \eqref{EQinternalForceField} can be further
systematically decomposed \cite{power,fortini14prl} into adiabatic
excess ($\fv_{\rm axc}$) and superadiabatic one-body contributions
($\fv_{\rm sup}$), according to
\begin{align}
  {\bf f}_{\rm int} = {\bf f}_{\rm axc} + {\bf f}_{\rm sup}.
  \label{EqSplitInt}
\end{align}
Here the excess (over ideal gas) adiabatic force field $\fv_{\rm axc}$
is the internal force field in a hypothetical equilibrium system which
has the same density profile as the real nonequilibrium system at time
$t$. 
Hence
\begin{align}
  {\bf f}_{\rm axc}(\rv,t) &= \rho(\rv,t)^{-1}
  \int d\rv^N \Psi_{{\rm ad},t}(\rv^N) \hat {\bf F}_{\rm int},
  \label{EQfadxAsAverage}
\end{align}
where the average is
over a canonical equilibrium distribution $\Psi_{{\rm ad},t}(\rv^N)$
for the (unchanged) interparticle interaction potential $u(\rv^N)$, but
under the influence of a hypothetical external (``adiabatic'')
one-body potential $V_{{\rm ad},t}(\rv)$, which is constructed in
order to yield in equilibrium the same one-body density as occurs in
the dynamical system at time~$t$~\cite{power,fortini14prl}, i.e.,
\begin{align}
  \rho(\rv,t)
  = \rho_{{\rm ad},t}(\rv)
  \equiv \int d\rv^N \Psi_{{\rm ad},t}(\rv^N) \hat\rho.
  \label{EQconditionForAdiabaticDensity}
\end{align}
The excess adiabatic force field is hence uniquely specified by
\eqref{EQfadxAsAverage} and \eqref{EQconditionForAdiabaticDensity};
computer simulations permit direct access \cite{fortini14prl}.  The
force splitting \eqref{EqSplitInt} then defines the superadiabatic
force field.

Here we demonstrate that $\fv_{\rm sup}(\rv,t)$ further splits
naturally and systematically into different contributions, which
correspond to different physical effects. We have shown before that
${\bf f}_{\rm sup}(\rv,t)$ contains viscous force contributions
\cite{delasHeras17gradvel}. These are of dissipative nature in that
they work against the colloidal motion (i.e.\ antiparallel to the flow
direction). Here we focus on the component of the superadiabatic force, which is
perpendicular to the local flow direction $\ev_v(\rv,t)$, where
$\vel=|\vel|\ev_v$; note that also $\Jv\parallel\ev_v$. We hence
define the (normal) structural force field $\fv_{\rm
  sup}^\perp(\rv,t)$ as the component perpendicular to the local flow
direction,
\begin{align}
  \fv_{\rm sup}^\perp &= \fv_{\rm sup} - \fv_{\rm sup}\cdot\ev_v\ev_v.
  \label{EQfsupNormal}
\end{align}
In contrast to the viscous force, the structural force is
nondissipative, since the associated power density vanishes
identically everywhere, $\Jv\cdot\fv_{\rm
  sup}^\perp=|\Jv|\ev_v\cdot\fv_{\rm sup}^\perp=0$.

The structural force plays a vital role in nonequilibrium, as it can
stabilize density gradients. In order to demonstrate this effect, we
consider a two-dimensional toy system of Gaussian core particles
\cite{GCMmodel} in an inhomogeneous external shear field. The pair
interaction potential is $\epsilon\exp(-r^2/\sigma^2)$, where $r$ is
the distance between both particles, and $\epsilon>0$ is the energy
cost at zero separation. We use $\epsilon$ and $\sigma$ as the energy
and the length scale, respectively. $N$ particles are located in a
square box of length $L$ with periodic boundary conditions and (unit
vector) directions $\ev_x$ and $\ev_y$ along the square box. The
driving occurs along ${\bf e}_x$ according to an inhomogeneous
external shear field,
\begin{align}
 \fv_{\rm ext}(y,t)= f_0 \sin(2\pi y/L)\theta(t){\bf e}_x,\label{EQfext}
\end{align}
where $f_0$ is a constant which controls the magnitude of the driving
force, and $\theta(\cdot)$ indicates the Heaviside (step) function,
such that the force is instantaneously switched on at time
$t=0$. Ultimately the system reaches a steady state with a density
gradient along ${\bf e}_y$, i.e.\ $\rho(\rv,\infty)=\rho(y)$, as we
will see below. The density gradient is then solely sustained by the structural force
$\fv_{\rm sup}^\perp$.

\begin{figure*}
\includegraphics[width=1.0\textwidth]{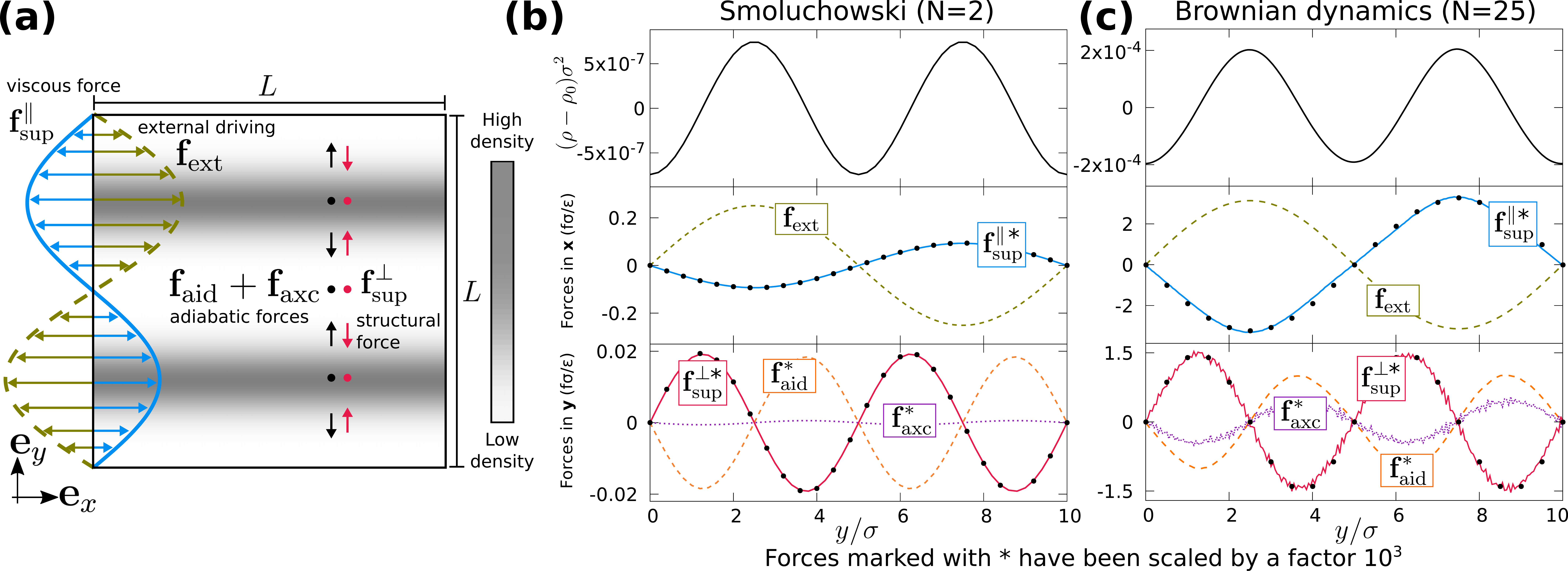}
\caption{(a) Illustration of the setup of a sinusoidal inhomogeneous
  shear flow induced by the external force field \eqref{EQfext}.
  Shown are the superadiabatic viscous force, which opposes the
  externally induced flow, the adiabatic excess and ideal forces,
  which tend to relax the density gradient, as well as the
  superadiabatic structural nonequilibrium force, which restores the
  force balance in the $y$-direction. (b) Steady-state density and
  force profiles obtained by numerically solving the Smoluchowski
  equation in a system with $N=2$, $L/\sigma=10$, and
  $k_BT/\epsilon=0.4$. (Top panel) Density profile for average density
  $\rho_0=N/L^2=0.02\sigma^{-2}$ as a function of $y$. (Middle panel)
  Forces acting along ${\bf e}_x$ as a function of $y$: external force
  $\fv_{\rm ext}$ of imposed amplitude $f_0\approx0.25\epsilon/\sigma$
  (green-dashed line), and superadiabatic viscous force (solid-blue
  line) $\fv_{\rm sup}^\parallel$. (Bottom panel) Forces acting along
  ${\bf e}_y$ as a function of $y$: adiabatic ideal (diffusion)
  $\fv_{\rm aid}$ (orange dashed line), adiabatic excess $\fv_{\rm
    axc}$ (violet dotted line), and superadiabatic structural force
  $\fv_{\rm sup}^\perp$ (solid-red line). (c) Same as (b) but for a
  system with $N=25$, $L/\sigma=10$, $\rho_0=0.25\sigma^{-2}$,
  $k_BT/\epsilon=1$ and $f_0\approx3.14\epsilon/\sigma$.  Data
  obtained using BD simulations. Forces marked with an asterisk in (b)
  and (c) have been multiplied by $10^3$ for clarity.  The black
  circles in panels (b) and (c) indicate the theoretical prediction
  \eqref{EQpredictions} of the superadiabatic viscous and structural
  forces.}
\label{fig1}
\end{figure*}

In order to study the system on the Fokker-Planck level, we solve numerically the exact many-body Smoluchowski equation (SE) for overdamped Brownian motion.
The time evolution of the probability distribution $\Psi$ is given exactly by the many body continuity equation
\begin{align}
\frac{\partial \Psi(\rv^N,t)}{\partial t}=-\sum_i\nabla_i\cdot\vel_i\Psi(\rv^N,t).\label{EQSmoluchowski}
\end{align}
where $\vel_i$ is the velocity of particle $i$, given by the force balance
\begin{align}
  \gamma \vel_i &= 
  - \nabla_i (u(\rv^N)+k_BT \ln \Psi(\rv^N,t))
  + {\bf f}_{\rm ext}(\rv_i,t).
  \label{EQofMotionI}
\end{align}
We solve \eqref{EQSmoluchowski} and \eqref{EQofMotionI} numerically
using a (standard) operator splitting
approach~\cite{NumericalRecipes}.  Each spatial coordinate is
discretized in increments $\Delta x=\sigma/5$; we use a time step
$\Delta t/\tau=5\times10^{-3}$ with timescale
$\tau=\sigma^2\gamma/\epsilon$. This method
provides exact results of the nonequilibrium dynamics up to numerical
inaccuracies. For $d$ space dimensions, the dimension of configuration
space is $Nd$, which limits the applicability of the method to systems
with small numbers of particles. Here, we consider $N=2$ in $d=2$,
which renders the numerical field $Nd=4$ dimensional; hence including
time, we solve a $4+1$ dimensional numerical problem.  As we show,
despite the limited number of particles, all relevant forces are
present.

In order to analyse larger systems, we use Brownian dynamics (BD),
i.e.\ integrating in time the Langevin equation of motion, which
corresponds to \eqref{EQSmoluchowski} and \eqref{EQofMotionI}:
\begin{align}
\gamma \frac{d\rv_i(t)}{dt}&=
 - \nabla_i u(\rv^N)
       + {\bf f}_{\rm ext}(\rv_i,t) + \boldsymbol\chi_i(t),
\label{EQBD} 
\end{align}
where $\boldsymbol\chi_i$ is a delta-correlated Gaussian random force.
We use a time step $dt/\tau=10^{-4}$ and histogram bins of size
$\Delta x=\sigma/20$. Density and force density profiles are obtained
by averaging over a total time of $\sim10^9\tau$ in steady state.

In both SE and BD we use the iterative scheme of
Refs.~\cite{fortini14prl,bernreuther16pre} to construct the adiabatic
external potential $V_{{\rm ad},t}(\rv)$.  The superadiabatic force
then follows immediately from~\eqref{EqSplitInt} since both ${\bf
  f}_{\rm int}$ and ${\bf f}_{\rm axc}$ can be directly calculated
(sampled) in SE (BD).  We have used the recently developed force
sampling method~\cite{delasHeras2017sampling} to improve the sampling
of the density profile in BD.  We expect, on principal grounds,
  that the SE and BD results agree (for identical values of system
  size and particle number, here $N=2$). 

A schematic showing all forces in steady state is shown in
Fig.~\ref{fig1}a.
The stationary density- and force-profiles obtained by solving the Smoluchowski equation $(N=2)$,
and using BD simulations $(N=25)$ are shown in Fig.~\ref{fig1}b and Fig.~\ref{fig1}c, respectively.
A net flow exists along $+{\bf e}_x$, since the external force is only partially balanced by a
superadiabatic force of viscous nature, $\fv_{\rm sup}^\parallel$, along  $-{\bf e}_x$ (see middle panels). Hence in this situation $\ev_v=\ev_x$.
The viscous force has roughly the same shape as the external force but with reversed direction.
The inhomogeneous external force~\eqref{EQfext}, $\fv_{\rm ext}(y)$, creates
a density gradient in ${\bf e}_y$ (see top panels), since the particles
migrate to the low shear rate regions.
The adiabatic ideal (i.e.\ diffusive) and adiabatic excess (i.e.\ due to internal interaction)
forces act along ${\bf e}_y$ and both try to relax the density gradient (see bottom panels). Both adiabatic
forces are, however, exactly balanced by a structural superadiabatic force $\fv_{\rm sup}^\perp$. The presence of the structural
force hence renders the inhomogeneities of the density profile stationary in time.

\begin{figure*}
\includegraphics[width=1.00\textwidth]{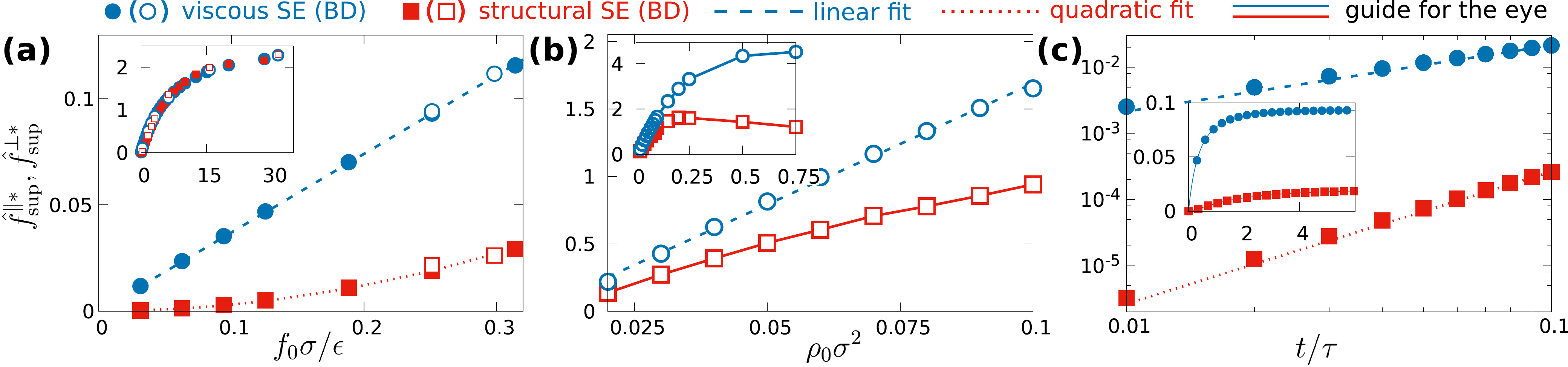}
\caption{Scaled amplitude of the superadiabatic forces $\hat f_{\rm
    sup}^{\alpha*}=10^3\hat f_{\rm sup}^{\alpha}\sigma/\epsilon$ with
  $\alpha\in\{\parallel,\perp\}$, as a function of: (a) the amplitude
  of the external driving force $f_0$, (b) the average density
  $\rho_0=N/L^2$, and (c) the time $t$. (a) Results for $L/\sigma=10$,
  $N=2$, and $k_BT/\epsilon=0.4$.  (b) Results in a system with
  $L/\sigma=10$, $f_0\approx3.14\epsilon/\sigma$, $k_BT/\epsilon=1$
  and varying $N$.  (c) Note the logarithmic scale in the main
  plot. The external driving force is switched on at $t=0$.  Results
  for $L/\sigma=10$, $N=2$, $k_BT/\epsilon=0.4$, and
  $f_0\approx0.25\epsilon/\sigma$.  In all panels, the viscous
  (structural) superadiabatic force is represented with blue circles
  (red squares), as indicated in the upper legend. Full (empty)
  symbols correspond to SE (BD) calculations.  The blue dashed (red
  dotted) line indicates a linear (quadratic) fit to the data, as
    detailed in the \cite{SupMat}. Solid lines are guides to the
  eye. The inset in each figure shows a linear plot of the same
  quantities as those in the main plot, but over an extended region.}
\label{fig2}
\end{figure*}

We next analyse the system more systematically by comparing the
behaviour of the structural force to that of the viscous force.  We
show in Fig.~\ref{fig2}a the amplitudes of the viscous force, $\hat
f_{\rm sup}^\parallel$, and the structural force, $\hat f_{\rm
  sup}^\perp$, (measured from maximum to baseline) as a function of
the amplitude of the external driving, $f_0$. For small driving the
viscous force scales linearly with $f_0$. This behaviour is expected,
since for weak driving the velocity is proportional to the strength of
the driving and the viscous force is proportional to the velocity
(Newtonian rheology).  The structural force, on the other hand, scales
quadratically with $f_0$ and hence also with the velocity in the small
driving regime.  Both forces saturate for high values of $f_0$, see
the inset in Fig.~\ref{fig2}a. Fig.~\ref{fig2}b shows $\hat f_{\rm
  sup}^\parallel$ and $\hat f_{\rm sup}^\perp$ as a function of the
average density $\rho_0$, revealing again profound differences between
viscous and structural forces. The viscous force increases linearly
with increasing $\rho_0$ at low densities and it saturates at high
densities. In contrast, the structural force is non-monotonic and
exhibits a maximum at an intermediate density.

We rationalize these findings by developing a theory within the power
functional approach~\cite{power}. Here the adiabatic and the
superadiabatic contributions to the internal force field
\eqref{EqSplitInt} are obtained via functional differentiation of two
generating functionals:
\begin{align}
  {\bf f}_{\rm axc}(\rv,t) &=
  -\nabla\frac{\delta F_{\rm exc}[\rho]}{\delta \rho(\rv,t)},
  \label{EQfadViaDerivative}\\
  {\bf f}_{\rm sup}(\rv,t) &=
  -\frac{\delta P_t^{\rm exc}[\rho,\Jv]}{\delta\Jv(\rv,t)}.
  \label{EQfsupViaDerivative}
\end{align}
Here $F_{\rm exc}[\rho]$ is the intrinsic excess (over ideal)
Helmholtz equilibrium free energy functional of density functional
theory, and $P_t^{\rm exc}[\rho,\Jv]$ is the excess (again over ideal)
superadiabatic functional of power functional theory \cite{power}.
Dynamical density functional theory (DDFT)
\cite{evans79,marinibettolomarconi99,archer04ddft} is obtained by
setting $P_t^{\rm exc}=0$; hence no superadiabatic forces (neither
viscous nor structural) occur in DDFT.  In
Ref.~\cite{delasHeras17gradvel}, using a change of variables in {power
  functional theory} from the current $\Jv$ to the velocity gradient
$\nabla\vel$, it is shown that the excess superadiabatic functional
can be represented as a functional of $\nabla\vel$. Following the same
line, we consider a temporally non-Markovian, but spatially local
form:
\begin{align}
  P_t^{\rm exc} &= \int d\rv \Big[\int_0^tdt'
    n_{tt'}
    (\nabla\times\vel)\cdot(\nabla\times\vel')
  \label{EQPtexcAll}  
  \\
  &\quad -\int_0^tdt'\int_0^tdt''m_{tt't''}
  (\nabla\cdot\vel)
  (\nabla\times\vel')\cdot(\nabla\times\vel'')\Big],\notag
\end{align}
where we have used the short-hand notation $\vel'=\vel(\rv,t')$ and
$\vel''=\vel(\rv,t'')$.
 The factors $n_{tt'}$ and $m_{tt't''}$ are density-dependent temporal
 convolution kernels; the subscripts indicate the time arguments; here
 the dependence is on the differences $t-t'$ and $t-t''$; the specific
 form of the kernels will depend on the form of the interparticle
 interaction potential $u(\rv^N)$.  In \eqref{EQPtexcAll} we have left
 away bi-linear and higher contributions in $\nabla\cdot\vel$
 \cite{delasHeras17gradvel}; these are important for compressional
 flow, but not for the present shear setup.
 Furthermore, as $\rho$ is practically constant in the cases
  considered, contributions in $\nabla\rho$ have also been omitted.

The superadiabatic force field is then obtained by inserting
\eqref{EQPtexcAll} into \eqref{EQfsupViaDerivative}, where the
derivative is carried out at fixed density, and hence $\delta/\delta
\Jv=\rho^{-1}\delta/\delta\vel$. Furthermore the spatial derivatives
in \eqref{EQPtexcAll} can be suitably re-arranged by (spatial)
integration by parts. The resulting force density is
\begin{align}
  \rho\fv_{\rm sup}(\rv,t) &= \int_0^tdt' 
  \nabla\cdot n_{tt'} \nabla\vel'
  \label{EQfsupWithMemory}\\
  &
\quad  -\int_0^tdt'\int_0^tdt''
    \nabla m_{tt't''}(\nabla\times\vel')\cdot(\nabla\times\vel'').
    \notag
\end{align}
In steady state and for the case of constant density profile,
\eqref{EQfsupWithMemory} reduces to:
\begin{align}
  \rho\fv_{\rm sup}(\rv) &= \eta \nabla^2 \vel 
  -\chi\nabla(\nabla\times\vel)^2,
  \label{EQfsupInSteadyState}
\end{align}
where the prefactors $\eta=\lim_{t\to\infty}\int_0^t dt'n_{tt'}$ and
$\chi=\lim_{t\to\infty}\int_0^tdt'\int_0^tdt''m_{tt't''}$ are moments
of the convolution kernels, which depend on the overall density. We
can identify $\eta$ with the shear viscosity
\cite{delasHeras17gradvel}, such that the first term in
\eqref{EQfsupInSteadyState} represents the viscous force. From
carrying out the normal projection \eqref{EQfsupNormal}, the second
contribution in \eqref{EQfsupInSteadyState} yields the structural
force $\fv_{\rm sup}^\perp(\rv,t)$.  Note that while the forms
  \eqref{EQfsupWithMemory} and \eqref{EQfsupInSteadyState} could
  possibly be postulated based on symmetry considerations alone, in
  the current framework, the generator \eqref{EQPtexcAll} constitutes
  the more fundamental object, as the form of the force field follows
  via the functional derivative \eqref{EQfsupViaDerivative}.
  Note that for the ideal gas ($u=0$) $P_t^{\rm exc}=0$ and $\fv_{\rm
    sup}$=0 by construction \cite{power}.

In accordance with our numerical results, we assume that the flow
field is dominated by the external force, hence we approximate
\eqref{EQofMotion} by $\vel(y,t)\approx\fv_{\rm
  ext}(y,t)/\gamma$.  Insertion of \eqref{EQfext} into
\eqref{EQfsupInSteadyState} then yields
\begin{align}
  \fv_{\rm sup}(y) &=
  -\frac{f_0\eta k^2}{\gamma\rho} \sin(ky)\ev_x
  +\frac{f_0^2\chi k^3}{\gamma\rho}\sin(2ky)\ev_y,\label{EQpredictions}
\end{align}
where $k=2\pi/L$. In Fig.~\ref{fig1}b and \ref{fig1}c we show the
comparison between the predicted (black circles) and the computed
superadiabatic forces using SE and BD (solid lines). The values of the
response coefficients $\eta$ and $\chi$, cf.~\eqref{EQpredictions},
have been adjusted to fit each amplitude, see 
 \cite{SupMat} for details.

The theory then predicts the shape of both viscous and
structural forces without further adjustable parameters, and it is in
excellent agreement with the results from SE and BD.  In addition, the
linear (quadratic) scaling of the viscous (structural) force with the
velocity, Fig.~\ref{fig2}a, is also accounted by the theory,
cf.~\eqref{EQfsupInSteadyState}. Due to the saturation of both
superadiabatic forces (inset of Fig.~\ref{fig2}a) it is necessary to 
analyze very small driving to obtain the correct scaling. 
Also, at low average density $\rho_0$
the viscous force is proportional to $\rho_0$ (see Fig.~\ref{fig2}b),
which according to \eqref{EQpredictions} implies
$\eta\propto\rho_0^2$, as expected
 \cite{reinhardt2013}.
 See \cite{SupMat} for results for strong driving conditions and
  for a theoretical prediction of the density profile.  

Memory plays a vital role in nonequilibrium systems, as we show in
Fig.~\ref{fig2}c by investigating the transient time evolution after
switching on the driving at $t=0$, cf.\ \eqref{EQfext}.  Both
superadiabatic force contributions vanish in equilibrium ($t\leq 0$)
and saturate in steady state ($t\rightarrow\infty$). At short times
after switching on the driving force, the viscous (structural) force
is linear (quadratic) in $t$, in full agreement with the non-Markovian
form of \eqref{EQPtexcAll}.  See 
  \cite{SupMat} for an analysis of the scaling of the amplitude of the
  forces with wavenumber~$k$. Here we have still taken $\ev_v=\ev_x$
as a (very good) approximation.

We conclude that the structural force is a primary candidate for a
universal mechanism that leads to nonequilibrium structure formation.
Examples of systems where a force acts perpendicular to the flow
include shear banding~\cite{dhont1999,dhont2003,dhont2014}, colloidal
lane formation~\cite{laningLitOne,laningLitTwo}, and effective
interactions in active spinning
particles~\cite{ActiveRotPRL,aragonesNC}.   The theory that we
  present here operates in a self-contained way on the one-body level
  of correlation functions, and hence is different from the approach
  of Refs.\ \cite{braderkrueger11molphys,scacchi2016}, where a dynamic
  closure on the two-body level via modelling of Brownian
  ``scattering'' is postulated.
 Note that the treatment of
  Refs.~\cite{braderkrueger11molphys,scacchi2016} relies on the
  density distribution as the fundamental variable; in contrast, our
  theory, predicts the behaviour of the system directly from the
  velocity field, cf.~\eqref{EQfsupWithMemory} and
  \eqref{EQfsupInSteadyState}.

 We have focused here on the viscous
and the structural contributions to the superadiabatic force in
shear-type flow. In compressional flow, where $\nabla\cdot\vel\neq 0$,
further force contributions can occur.  Calculating the values for
  the transport coefficients $\eta$ and $\chi$ within the current
  theory is desirable, based, e.g., on the two-body level of
  correlation functions \cite{NOZ}. 
 Furthermore, interesting
  research task for the the future include investigating the effects
  of inhomogeneous temperature fields \cite{falasco2016}, and the
  possible emergence of time-periodic states (``time
  crystals''\cite{timeCrystals}) in inertial dynamics
  \cite{inertialPFT}.

This work is supported by the German Research Foundation (DFG) via
SCHM 2632/1-1.  N.~C.~X.~S. and T.~E. contributed equally to this
work.

\section{Supplemental Material: Structural nonequilibrium forces in driven colloidal systems}

\subsection{Superadiabatic force profiles under strong driving conditions}

The examples presented in the main text address the behaviour of the
system in response to relatively weak external perturbations.  Here we
show that under strong external driving conditions, both
superadiabatic force profiles change significantly. We show in
Supplemental Fig.~\ref{figS1} the viscous and the structural
superadiabatic forces for both weak and for strong external driving
conditions. We find that the direction of both force fields remains
unchanged upon increasing the driving. Hence, for the cases considered
here, the viscous force opposes always the external force and the
structural force sustains (counteracts) the density gradient. However,
although the only difference between both driving conditions is the
magnitude of the external driving, the superadiabatic
(position-dependent) force profiles change significantly their shape
depending on the strength of the driving. 
The power functional approximation presented in Eq.~(15) of the main text
is intended to deal with low/moderate driving and does
not fully reproduce the characteristics of the superadiabatic forces
at high driving conditions. Note that the magnitude of the superadiabatic force is
relatively small as compared to the magnitude of the external force,
even for the case of weak driving (Fig.~1 of the main
text). Therefore, the velocity field basically follows the external
driving, and the superadiabatic forces are then given by Eq.~(18) of
the main text. That is, the theory does not reproduce the shape change
of the superadiabatic forces with the magnitude of the external
driving (Supplemental Fig.~\ref{figS1}).  This was expected since in
Eq.~(15) we have only incorporated the first two terms of the
expansion of $P_t^{\text{exc}}$ in powers of $\nabla\vel$. Therefore,
one expects Eq.~(15) to be valid under weak driving conditions. The
correct treatment of strong driving conditions requires the addition
of higher order terms to Eq.~(15).  It is straightforward to
incorporate such terms by rewriting the excess superadiabatic
functional as

\begin{align}
  P^{\rm exc} &= \int d\rv \Big[\eta
    (\nabla\!\times\!\vel)^2-\chi 
  (\nabla\!\cdot\!\vel)(\nabla\!\times\!\vel)^2\Big]h(\nabla\vel),\label{EQpexcNonlinear}
\end{align}
where $h(\nabla\vel)$ is a scalar-valued function, which we choose to
have the following form:
\begin{equation}
h(\nabla\vel)=\frac{1}{1+c_0 (\nabla\times\vel)^2},
\end{equation}
where $c_0$ is a positive constant which controls the strength of the
influence of the flow; setting $c_0=0$ recovers the form of Eq.~(15)
of the main text. In Eq.~\eqref{EQpexcNonlinear} we have assumed
steady state conditions and therefore eliminated the time
integrals. The first two terms of a series expansion of
Eq.~\eqref{EQpexcNonlinear} in $\nabla\vel$ have the same functional
structure as that presented in Eq.~(15) of the main text. In addition,
Eq.~\eqref{EQpexcNonlinear} incorporates higher order contributions in
$\nabla\vel$. We find that the superadiabatic force profile that
result upon functional differentiation of Eq.~\eqref{EQpexcNonlinear}
is in excellent agreement with the simulation results both for weak
and for strong driving conditions (see black circles in Supplemental
Fig.~\ref{figS1}). For both cases shown in the figure, we have set
$c_0=3.5\cdot10^{-3}\;\;\tau^2$ and fit the values of $\eta$ and
$\chi$. The values of $\eta$ and of $\chi$ determine the overall
amplitude of the viscous and of the structural superadiabatic force,
respectively. We find that the values of $\eta$ and $\chi$ seem to
depend on the magnitude of the external driving, which is a clear
indication that Eq.~\eqref{EQpexcNonlinear} constitutes only a first
approximation.

\begin{figure*}
\centering
\includegraphics[width=0.85\textwidth]{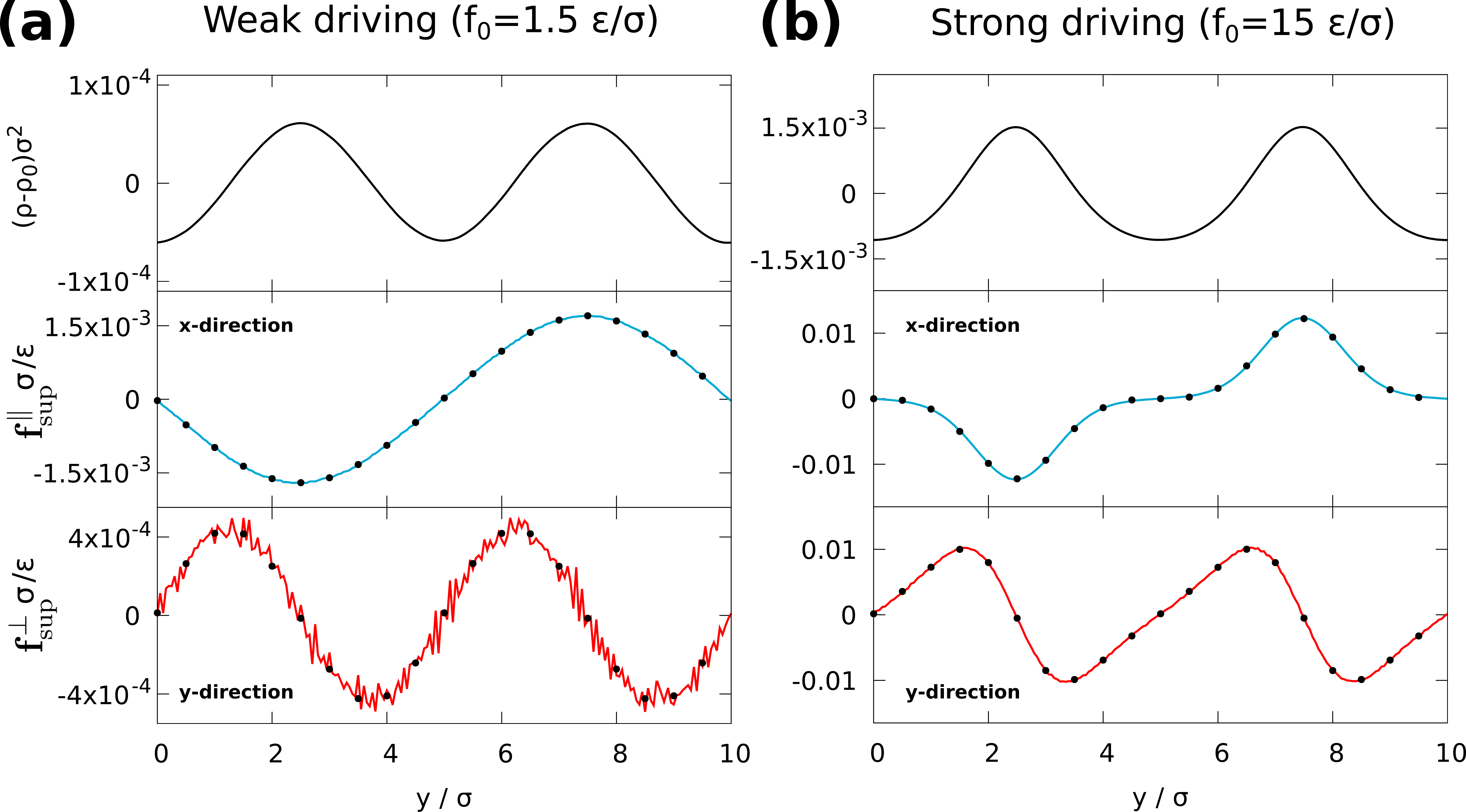}
\caption{
  Steady-state density and force profiles according to BD simulations
  in a system with $N=25$, $L/\sigma=10$, and $k_BT/\epsilon=1.0$.
  Results are shown for two strengths of the external driving force: (a) 
  $f_0\approx1.5\epsilon/\sigma$, weak external driving, and (b) 
  $f_0\approx15\epsilon/\sigma$, strong external driving. 
  (Top panels) Density profile, measured from the average density
  $\rho_0=N/L^2=0.25\sigma^{-2}$, as a function of $y$. (Middle panel)
  Superadiabatic viscous force acting along ${\bf e}_x$ as a function of $y$
  obtained by BD simulations (solid-blue line) and theoretically (black circles).
  (Bottom panel) Superadiabatic structural force acting along
  ${\bf e}_y$ as a function of $y$ obtained by BD simulations 
  (solid-red line) and theoretically (black circles). 
  }
\label{figS1}
\end{figure*}

\subsection{Scaling of the superadiabatic forces with the inverse box length}
We present a scaling analysis of the magnitude of the viscous and of
the structural superadiabatic force fields with the amplitude of the
external driving and the overall density in the main text. Here, we
focus on the scaling of these force fields with the wavenumber
$k=2\pi/L$ that characterizes the inhomogeneous shear profile,
cf.~Eq.~(9) of the main text. In practice we change $k$ by changing
the box size $L$, ensuring that the other relevant variables are
unchanged, e.g.\ keeping the mean density constant by accordingly
modifying the number of particles, $N$.

Our theory, cf.~Eq.~(18) of the main text, predicts different types of
scaling with $k$ for the viscous (quadratic in $k$) and for the
structural (cubic in $k$) superadiabatic forces. We fix the amplitude
of the external driving ($f_0\approx0.30\epsilon/\sigma$) and the
average density of the system ($\rho_0\sigma^2=0.02$). Therefore,
according to Eq.~(18) of the main text, the only remaining dependence
of the amplitude of the superadiabatic forces is on $k$. 
In Supplemental Fig.~\ref{figS2} we show the amplitude of the
superadiabatic forces as a function of the wavenumber. The smallest
system corresponds to $N=2$ and $H/\sigma=10$, and the largest one to
$N=100$ and $H/\sigma=70.7$. Each curve shows three different
regimes. For small boxes (large wavenumber) we observe a saturation of
the superadiabatic forces, similar to the saturation effect observed
by increasing the external driving or the density (see Fig.~2 of the
main text).  At intermediate wavenumbers the forces scale linearly
with $k$. Finally, for long boxes (small wavenumbers) the curves
depart significantly from a linear behaviour. Note that both curves
must pass through the origin of coordinates. The low driving
conditions that are implicit in going to small values of $k$ render
the analysis of the data difficult. Nevertheless, the scaling of the
viscous force is fully consistent with the quadratic prediction of the
theory, Eq.~(18) of the main text. Also, it is evident from the data
that the structural force is not linear in $k$ for sufficiently large
boxes (small wavenumbers). The precision of our data, however, does
not allow us to confirm the cubic scaling predicted by the theory,
although the data is also clearly not in contradiction to the
analytical result. We hence cannot rule out the contribution of
further terms to $P_t^{\rm exc}$. In particular, a term involving the
combination $(\nabla\cdot\vel)(\vel\cdot\vel)$ produces a structural
force with the same shape, but a linear scaling with $k$.  We conclude
that revealing the true scaling of the superadiabatic forces is a
major challenge. Due to the saturation effect, the forces increase
linearly in a broad range of system sizes. Only for very small
perturbations (very large system sizes in this case) the true scaling
behaviour is revealed.

\begin{figure}
\centering
\includegraphics[width=0.95\columnwidth]{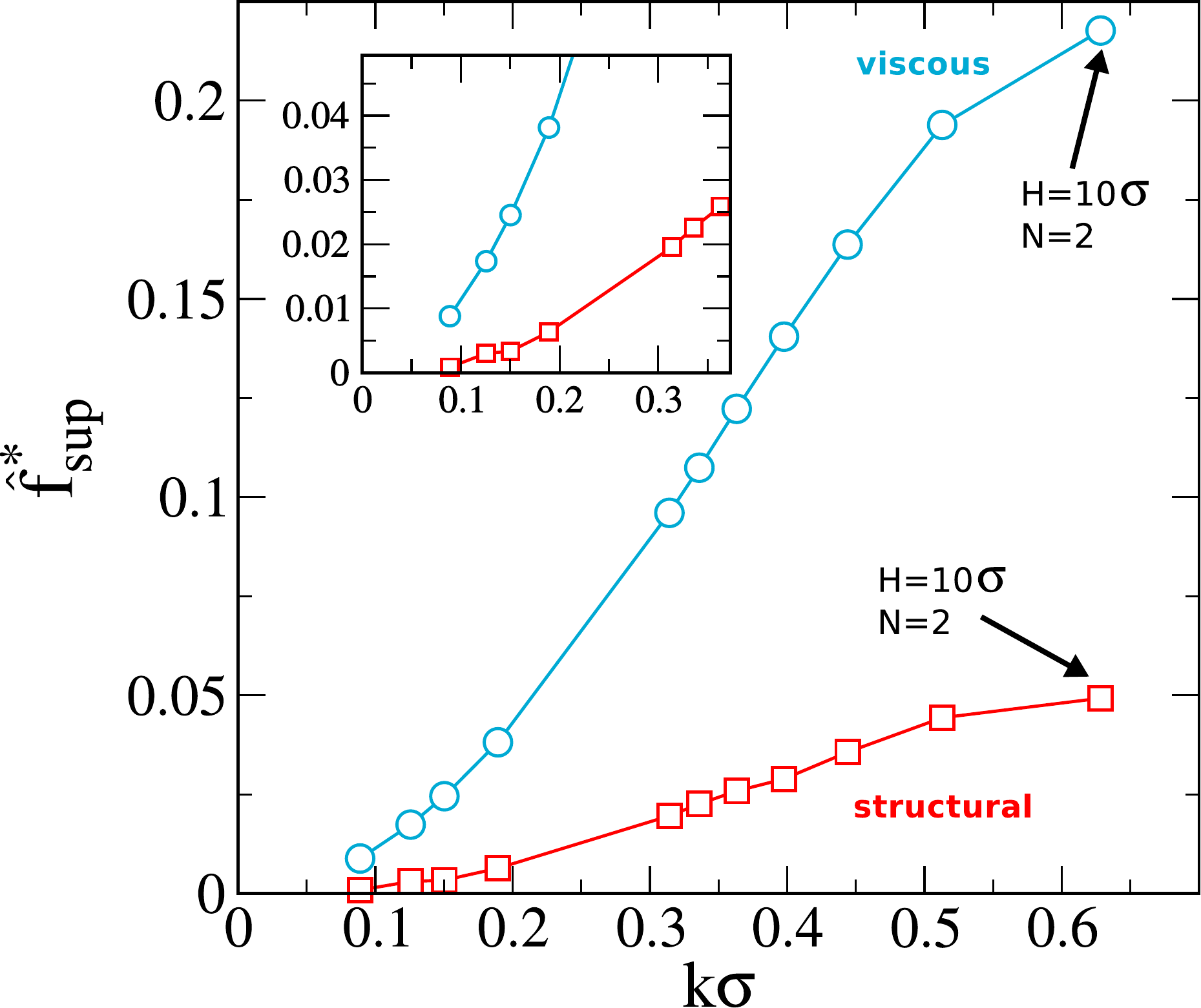}
\caption{ Scaled amplitude of the superadiabatic forces $\hat f_{\rm
    sup}^{*}=10^3\hat f_{\rm sup}^{}\sigma/\epsilon$ as a function of
  the (scaled) wavenumber $k\sigma=2\pi\sigma/L$. The blue circles
  (red squares) correspond to the viscous (structural) superadiabatic
  force, as indicated.  The lines are guides to the eye. Data obtained
  with BD simulations for systems with $k_BT/\epsilon=0.4$ and
  $f_0\approx0.30\epsilon/\sigma$.  The number of particles and the
  length of the box vary across the data points, such that the average
  density $\rho_0\sigma^2=0.02$ is kept constant. Both curves must
  pass through the origin.  }
\label{figS2}
\end{figure}

\subsection{Representation of numerical results via fit functions}

The fit parameters in Fig.~2a of the main text are: $A_1 = 3.7 \cdot
10^{-4}$ (for the viscous force, linear in velocity), $A_2 = 3.0 \cdot
10^{-5}\sigma/\epsilon$ (for the structural force, quadratic in velocity). The fit
has been done in the region of weak driving $f_0\sigma/\epsilon\lesssim0.3$.

Regarding the scaling of the superadiabatic forces with the total 
density, Fig.~2b of the main text ($f_0\approx3.14\epsilon/\sigma)$, a good numerical representation 
in the range $0<\rho_0\sigma^2<0.1$ is given by $\eta(\rho_0)\approx1.7\cdot10^{-2}\rho_0^2\epsilon\tau\sigma^2$ and 
$\chi(\rho_0)\approx 10^{-2}\rho_0^2\epsilon\tau^2\sigma^2 - 0.4\cdot10^{-2}\rho_0^3\epsilon\tau^2\sigma^4$.

We reemphasize that the true scaling of viscous and structural
forces is only revealed at weak driving conditions and low densities.
At intermediate driving and/or total density both forces are linear 
in velocity and total density due to a, yet to be investigated,
saturation mechanism of the superadiabatic forces.

\subsection{Predicting the density profile}
We can obtain a prediction of the density profile from the
  theory, by projecting the force balance equation (Eq. (1) of the main text) onto
  $\ev_y$ and observing that $\vel\cdot\ev_y=0=\fv_{\rm
    ext}\cdot\ev_y$. We obtain $k_BT \partial \rho/\partial y = f_{\rm
    sup,y}$ upon neglecting the adiabatic excess contribution
  $\fv_{\rm adx}$, as this is small against the ideal term, $|\fv_{\rm
    adx}|\ll|k_BT\nabla\rho|$ in the low density case
  $N=2$. Integrating in $y$ gives to lowest order in the difference
  $\rho-\rho_0$ the result $\rho(y) = \rho_0-f_0^2\chi
  k^2\cos(2ky)/(2\gamma k_BT)$, which describes the low-density case
  (Fig.~\ref{figS1}b of the main text) quite well.

\end{document}